\documentclass[12pt]{iopart}
\usepackage{iopams} 

\newcommand{\algn}[2]{\begin{eqnarray} {#1} \label{#2} \end{eqnarray}}
\newcommand{\bra}[1]{\langle\hspace{0.45pt}{#1}\vert}
\newcommand{\ket}[1]{\vert{#1}\hspace{0.45pt}\rangle}
\newcommand{\calH}{\mathcal{H}}
\newcommand{\vv}[1]{\boldsymbol #1}

\begin{document}

\title{Reconstruction of Quantum Mechanics with Information Operators}

\author{Ken'ichi Takano}

\address{Toyota Technological Institute, 
Tenpaku-ku, Nagoya 468-8511, Japan}

\begin{abstract}
We reconstruct quantum mechanics by introducing 
"information operators" and excluding the concept of 
wave functions. 
Multiple information operators simultaneously describe 
a single system and continuously develop in time 
even in the process of a measurement. 
We also introduce the concept of condensation for a system 
with many degrees of freedom in a rather general meaning. 
In terms of the multiplicity of description and the condensation, 
we explain quantum phenomena including measurements 
without the collapse of the wave function. 
\end{abstract}

\pacs{03.65.-w, 03.65.Ta, 03.65.Ca}

\section{Introduction}

Quantum mechanics has been developed continuously 
from the discovery by Heisenberg and Schr\"{o}dinger. 
The principles of quantum mechanics are clearly mentioned 
in the traditional book of Dirac \cite{dirac}. 
The development includes improved formalisms and 
various calculation methods, and produces many applications. 
Nowadays, by a highly developed experimental technology, 
it is even possible to observe the interference effect of 
a single electron, which is proper to 
quantum mechanics \cite{tonomura}. 
Further it is now a realistic destination to control and utilize 
fundamental quantum effects like quantum entanglement. 

On the other hand, the interpretation of quantum mechanics 
has been controversial from the time of discovery. 
In particular, researchers have argued the physical meaning 
of a wave function from various standpoints. 
Among them, the Copenhagen interpretation is the most 
accepted one. 
However, there are many people who do not convince themselves that 
the interpretation explains all the quantum phenomena. 

In the interpretation issue, the collapse of the wave function is 
the most serious one. 
The wave function describing a physical system 
seems to change discontinuously 
from a time before the measurement to a time after it. 
Following von Neumann~\cite{neumann}, if we pursue 
this problem, it comes at the human consciousness. 
I think that this argument is inevitable if we consider 
the problem within the conventional framework of 
quantum mechanics. 
In my opinion, von Neumann has not completed the argument, 
since he did not argue what the collapse of a wave function 
at the consciousness is in detail. 

The existence of a mixed state brings further difficulties 
to the interpretation of quantum mechanics. 
While a pure state is represented as a wave function, 
a mixed state can only be represented as a density operator. 
Without the mixed state, 
we cannot describe many quantum 
phenomena in real experiments. 
Despite its necessity, there is unclearness in the meaning of 
the mixed state. 

In a typical explanation, the mixed state is not a state, and 
the system must be strictly in a single pure state 
described by a state vector, or a wave function. 
However, if we have not enough knowledge of the state vector, then we 
express the situation with lacked information as a mixed state 
described by a density operator. 
In the explanation, the wave function is fundamental 
and the density operator is secondarily constructed 
of wave functions. 
This idea seems to be reasonable, if we statistically treat 
an ensemble consisting of many equivalent systems where 
each system is strictly in one of the possible state vectors. 
By this interpretation, we can calculate the average 
of a physical quantity over the ensemble without ambiguity. 

The above argument is not applicable, if we treat a single quantum 
system and perform a single measurement to it. 
In this situation, there is no ensemble and we do not repeat the same 
experiment.  
Hence I doubt the proposition that the density operator is 
a secondary concept and what is essential is a state vector. 
Furthermore, the discontinuity corresponding to the collapse of 
a wave function still remains irrespective of the interpretation 
of the mixed state. 
We lack substantial understanding of both 
the pure state and the mixed state. 

In this paper, we reconstruct quantum mechanics 
on a standpoint totally different from the conventional one. 
In the reconstructed quantum mechanics, 
we have no wave functions. 
Instead, we introduce 'information operators' which carry 
information on the system. 
Although an information operator is mathematically similar to 
a density operator of the conventional quantum mechanics, 
it is physically different. 
In particular, there exist multiple information operators which 
simultaneously describe the same system. 
When we have an information operator, 
the reconstructed quantum mechanics gives 
an information operator at a later time, 
which may be a time after a measurement. 
Another essential concept is the condensation of a system. 
We show that a condensed system plays a role of 
an apparatus of measurement. 
By the concepts of the multiple description and the condensation, 
we consistently explain quantum phenomena 
including measurements 
without discontinuity like the collapse of a wave function. 

This paper is organized as follows. 
In section 2, we examine multiple descriptions of a classical dice 
for an instructive preparation. 
In section 3, we introduce information operators to 
describe multiply a quantum system. 
In section 4, we mention the time-development of 
the information operator. 
In section 5, we introduce the concept of condensation 
for a system with large degrees of freedom. 
In section 6, we define the composite system. 
In section 7, we argue the condensation of a composite system. 
In section 8, we define the measurement and the observable. 
In section 9, we introduce the information vector, 
which corresponds to the wave function in the conventional 
quantum mechanics. 
In section 10, we demonstrate the reconstructed quantum mechanics 
with the interpretation of the Stern-Gerlach experiment. 
In section 11, we examine an electron passing through the slits 
in a screen in terms of the reconstructed quantum mechanics. 
Section 12 is devoted to summary.

\section{Multiple Description of a Dice}

We treat multiple inequivalent descriptions of a single system 
in the following sections. 
To prepare for this unconventional idea, 
we examine multiple descriptions of a usual dice, 
although it is a classical object and 
the analogy is rather restrictive. 

When we roll a dice and have the 3-spot, 
the following four descriptions are all correct: 

(i) The dice shows any number of spots; 
$A=\{ 1, 2, 3, 4, 5, 6 \}$. 

(ii) The dice shows odd spots; 
$B=\{ 1, 3, 5 \}$. 

(iii) The dice shows less than or equal to 3 spots; 
$C=\{ 1, 2, 3 \}$. 

(iv) The dice shows just three spots; 
$D=\{ 3 \}$. 

Here each description is followed by the corresponding set. 
We say, for example in (ii), that the dice is described by $B$. 
We also say that $A$ is the maximum description and $D$ is 
the minimum description of the dice. 
On the other hand, the following descriptions are incorrect: 

(v) The dice shows even spots; 
$F=\{ 2, 4, 6 \}$. 

(iv) The dice shows just one spot; 
$G=\{ 1 \}$. 

Hence the dice is neither described by $F$ nor $G$. 

Now suppose we only have the information 
that the dice is described by $B$. 
In other words, we only know that the dice shows 
one of the 1, 3 and 5-spots. 
Then we ask which set describes the dice except for $B$. 
For example, the dice is described by $A$, too, since 
the cases of the 1, 3 and 5-spots are included 
in the cases of the 1, 2, 3, 4, 5 and 6-spots. 
In general, if $X \supset Y$, then the following is correct: 
if the dice is described by $Y$, then it is also described by $X$. 
Since $A$ is the maximum description, it always describes the dice. 

Next, we consider whether the dice is described by $D$ or not, 
when the dice is described by $B$. 
Since the dice may show one of the cases of the 1, 3 and 5-spots, 
we cannot definitely say that the dice is described by $D=\{ 3 \}$. 
It might be described by $G=\{ 1 \}$. 
In general, if $X \supset Y$, then the following is correct: 
if the dice is described by $X$, then it is not necessarily 
described by $Y$. 
Instead, we say that the dice is described by $D$ 
with probability $\frac{1}{3}$, since the possibility is 
one of the three sets, $\{ 1 \}$, $\{ 3 \}$ and $\{ 5 \}$. 

Further, we compare $B$ with $C$. 
There is no inclusion relation between them, 
although both describe the dice with the 3-spot simultaneously. 
This is an example that two sets simultaneously describe 
the dice without no inclusion relation between them. 

In the quantum mechanics reconstructed below, 
a system is also multiply described. 
However the description is crucially different from that of a dice. 
We have described the dice with the 3-spot 
by $A$, $B$ or $C$ because of the lack of information. 
In contrast, for a system in the reconstructed quantum mechanics, 
there is no idea corresponding to the fact that the dice actually shows the 3-spot. 
What we can ask is only about the information given or obtained. 
In this way, we will avoid the discontinuity in the observation 
of the quantum mechanics. 
Another fundamental difference is that a system is described by 
operators instead of sets.

\section{Information Operators}

We reconstruct quantum mechanics 
starting from the following postulate: 

\medskip
\noindent
{\bf Postulate 1 (Existence of a Hilbert space):} 
{\it There is a complex Hilbert space $\calH_S$ for a system $S$.} 
\medskip

In this Hilbert space $\calH_S$, we define operators 
which can represent information of the system $S$, as follows: 

\medskip
\noindent
{\bf Definition 1 (Information operator):} 
{\it When an Hermitian operator $\rho$ in the Hilbert space $\calH_S$ 
satisfies the condition, 
\algn{\mathrm{tr}\{\rho \} = 1 , \quad \rho \ge 0,}{info_operator}
it is an information operator, or an i-operator shortly.} 
\medskip

When the dimension of $\calH_S$ is of a finite value $d$, 
we call $\rho^{\rm max} \equiv \frac{1}{d}I$ 
the maximum i-operator, where $I$ is the identity operator. 
If the dimension of $\calH_S$ is infinite, 
there is no maximum {i-operator}. 
An i-operator $\rho^{\rm pure}=\ket{\psi}\bra{\psi}$ with 
 any normalized vector $\ket{\psi}$ in $\calH_S$ is called 
a pure i-operator. 

Any i-operator is represented as 
\algn{\rho =\sum_i \, p_i \, \ket{\psi_i} \bra{\psi_i}}{rho_sum}
with a set of appropriate normalized vectors $\{ \ket{\psi_i} \}$ 
in $\calH_S$, where $p_i  \ge 0$ for all $i$ 
and $\sum_i \, p_i = 1$. 

The following is a postulate about the description of a system. 

\medskip
\noindent
{\bf Postulate 2 (Description of the system):} 
{\it System $S$ is described by i-operators in $\calH_S$.} 
\medskip

This means that we can derive some information from 
the i-operator about the system $S$. 
In other words, 
we can predict results of a measurement with probabilities 
which the i-operator gives, as will be argued later. 
It is an important part for solving a problem to find an i-operator 
which describes the system under consideration. 

Postulate 2 never means that an i-operator uniquely describes 
the system, and, in general, there are multiple i-operators 
simultaneously describing the same system. 
If one knows more than two i-operators to describe a system, 
one can choose anyone of them as will be explained later. 

In the postulates of this paper, we do not require a {\it state}, 
which plays a central role in the conventional quantum mechanics. 
It is an incorrect idea that an i-operator incompletely 
describes a system owing to the lack of information. 
Multiple i-operators describing the same system equally satisfy 
the same postulates in this paper.  
In particular there is no absolute i-operator 
which plays a special role. 

Here we mention a definition and a postulate 
about a special relation between two i-operators. 

\medskip
\noindent
{\bf Definition 2 (Expansion and contraction of an i-operator):} 
{\it If there exists an operator $K$, and 
i-operators $\rho_1$ and ${\rho_2}$ satisfy the following relation, 
\algn{K \rho_1 K^{\dagger} = \rho_2,}{rho_expansion}
then $\rho_1$ is an expansion of $\rho_2$, and 
$\rho_2$ is a contraction of $\rho_1$. 
$K$ is the contracting operator which contracts $\rho_1$ 
to ${\rho_2}$.} 

\medskip
\noindent
{\bf Postulate 3 (Description of the system by an expansion):} 
{\it If an i-operator describes a system, 
any expansion of the i-operator also describes the system.}
\medskip

As for the reverse, even if an i-operator $\rho$ describes a system, 
$K \rho K^{\dagger}$ with any operator $K$ 
does not necessarily describe the system. 
We have the following theorems about the multiple descriptions 
by i-operators. 

\medskip
\noindent
{\bf Theorem 1 (Unitary transformation):} 
{\it If an i-operator describes a system, then another i-operator 
unitarily transformed from it also describes the system.} 

\medskip
\noindent
{\bf Proof:} 
In (\ref{rho_expansion}), 
the contracting operator $K$ may be a unitary operator. 
\hfill $\square$

\medskip
\noindent
{\bf Theorem 2 (Description by the maximum i-operator):} 
{\it If the dimension of the Hilbert space of a system is finite, 
the maximum i-operator always describes the system.} 

\medskip
\noindent
{\bf Proof:} 
We show that the maximum i-operator is an expansion of 
any i-operator. 
Let $\rho_1 = \frac{1}{d}I$ in (\ref{rho_expansion}) 
so that $K K^{\dagger} = d \, \rho_2$. 
For arbitrary $\rho_2$, we take the representation 
where $\rho_2$ is diagonal and its eigenvalues are 
$\rho_2^{(i)}$ ($i=1, 2, \cdots , d$). 
We take $K$ such that the diagonal elements 
are $\sqrt{\rho_2^{(i)}d}$ ($i=1, 2, \cdots , d$) and 
all the off-diagonal elements are 0. 
Then $K K^{\dagger} = d \, \rho_2$ is satisfied. 
\hfill $\square$

\medskip
\noindent
{\bf Theorem 3 (Composition of i-operators):} 
{\it If i-operator $\rho$ is composed of 
i-operators $\rho_i$ $(i = 1, 2, \cdots)$, i.~e. 
\algn{\rho = \sum_i p_i \rho_i}{composition}
with $\sum_i p_i = 1$ and $p_i > 0$ $(i = 1, 2, \cdots)$, 
then each $\rho_i$ is a contraction of $\rho$.} 

\medskip
\noindent
{\bf Proof:} 
Let the eigenvalues of $\rho$ be $\rho^{(1)}$, $\rho^{(2)}$, 
$\cdots$ in ascending order, and the normalized eigenvectors 
belonging to them be $\ket{\rho^{(1)}}$, $\ket{\rho^{(2)}}$ 
$\cdots$. 
Similarly let the eigenvalues of $\rho$ be $\rho_1^{(1)}$, 
$\rho_1^{(2)}$, $\cdots$ in ascending order, and 
the normalized eigenvectors belonging to them be 
$\ket{\rho_1^{(1)}}$, $\ket{\rho_1^{(2)}}$ $\cdots$. 
Clearly, the number of zero eigenvalues of $\rho$ is 
smaller than that of $\rho_1$. 
Hence there exist non-negative real numbers 
$a_i$ $(i = 1, 2, \cdots)$ which satisfy 
$a_i \rho^{(i)} = \rho_1^{(i)}$ $(i = 1, 2, \cdots)$. 
By taking $K$ as 
$K \equiv \sum_i \sqrt{a_i} \ket{\rho_1^{(i)}} \bra{\rho^{(i)}}$ 
we have $K \rho K^{\dagger} = \rho_1$. 
\hfill $\square$

\medskip

We define the amount of information of an i-operator by 
the entropy below: 

\medskip
\noindent
{\bf Definition 3  (Entropy):} 
{\it The amount of information, or the entropy of 
an i-operator $\rho$ is defined as} 
\algn{E[\rho] = - \mathrm{tr}\{\rho \log \rho \}.}{information}
\medskip

In particular, we have $E[\rho^{\rm max}] = \log d$ and 
$E[\rho^{\rm pure}] = 0$. 
The entropy of a general i-operator is a value between these. 
It is emphasized that the entropy is defined for each i-operator 
describing a system and not for the system itself. 
Different i-operators may have different entropies 
even for the same system. 

Below we mention a postulate about the idea that 
a system is described by an i-operator with probability. 

\medskip
\noindent
{\bf Postulate 4 (Description with probability):} 
{\it Let an i-operator $\rho$ be composed of i-operators 
$\rho_1$ and $\rho_2$ as 
\algn{\rho = p_1 \rho_1 + p_2 \rho_2}{probability}
with $p_1 > 0$, $p_2 > 0$ and $p_1 + p_2 = 1$. 
If $\rho$ describes a system, 
then $\rho_1$ describes the  system with probability $p_1$, 
and $\rho_2$ does with probability $p_2$.} 
\medskip

In Postulate 4, $\rho_1 \rho_2 = 0$ is not required. 
If we write (\ref{probability}) as 
$\rho = \sigma_1 + \sigma_2$, then we have the probabilities, 
$p_1 = \mathrm{tr} \, \sigma_1$ and 
$p_2 = \mathrm{tr} \, \sigma_2$. 
By using (\ref{probability}) successively, Postulate 4 is 
applicable to the case of an i-operator composed of 
more than two i-operators $\rho_i$ $(i= 1, 2, \cdots )$: 
i.~e. if $\rho = \sum_i p_i \rho_i$ 
describes a system, then the probability that 
$\rho_i$ describes the system is $p_i$. 

For example, we consider a system which consists only of 
a spin $\vv{s} = (s_x, s_y, s_z)$ of magnitude 1. 
Let the eigenvectors of $s_z$ belonging to 
eigenvalues $+1$, 0 and $-1$ be 
$\ket{+}$, $\ket{0}$ and $\ket{-}$, respectively. 
When we have information that the present eigenvalue 
of the system is not zero, we seek an i-operator 
describing the system. 
An i-operator describing the system is clearly 
$\rho' = \frac{1}{2} (\ket{+}\bra{+} + \ket{-}\bra{-})$, 
since the eigenvalue of $s_z$ is not zero, and it is $+1$ or $-1$ 
with equal possibility. 
By Theorem 2, i-operator $\rho^{\mathrm{max}}$ = 
$\frac{1}{3} (\ket{0}\bra{0} + \ket{+}\bra{+} + \ket{-}\bra{-})$
also describes the system. 
There is no contradiction in that multiple i-operators simultaneously 
describe the same system. 
$\rho'$ describes the system more definitively than 
$\rho^{\mathrm{max}}$ does, 
because $E[\rho'] = \log 3$ and 
$E[\rho^{\mathrm{max}}] = \log 2$. 
We next consider i-operators $\rho^{+} = \ket{+}\bra{+}$ 
and $\rho^{-} = \ket{-}\bra{-}$ in this case. 
Both $\rho^{+}$ and $\rho^{-}$ are 
contractions of $\rho'$ from Theorem 3, 
since $\rho'$ = $\frac{1}{2} (\rho^{+} + \rho^{-})$. 
Then they do not describe the system definitively. 
From Postulate 4, the system 
is described by $\rho^{+}$ with probability $\frac{1}{2}$, and 
is described by $\rho^{-}$ with probability $\frac{1}{2}$. 

\section{Time Development}

We consider the time development of the i-operator as follows. 

\medskip
\noindent
{\bf Postulate 5 (Time development of the i-operator):} 
{\it Let i-operator $\rho(t)$ describe system $S$ at time $t$. 
If the system $S$ is isolated until time $t'$ after $t$, then 
there exists a unitary operator $U(t, t')$ proper to the system 
such that the i-operator 
\algn{\rho(t') = U(t, t') \rho(t) \, U^{\dagger}(t, t')}{time_unitary}
describes the system at time $t'$. 
$\rho(t)$ is continuous with respect to $t$.} 
\medskip

If we have an i-operator describing a system at time $t$, 
we can calculate an i-operator describing it at a later time $t'$ 
from (\ref{time_unitary}) until the system is isolated. 
Unitary operator $U(t, t')$ is proper to the system, and 
any i-operators describing the same system develop 
by the same $U(t, t')$. 
$U(t, t')$ is continuous with respect to $t$ and $t'$ 
since $\rho(t)$ is continuous. 
As will be mentioned later, a time development 
with measurement is also continuous. 

Under the time uniformity, the unitary operator $U(t, t')$ 
is written with an Hermitian operator $H$ as 
\algn{U(t, t') = \exp \{-i(t'-t)H/\hbar  \},}{time_operator}
where $\hbar$ is Plank's constant divided by $2\pi$. 
Then we call $H$ the Hamiltonian. 

From (\ref{time_unitary}) and (\ref{time_operator}), 
an i-operator $\rho(t)$ describing the system satisfies 
\algn{i\hbar \frac{d\rho}{dt} = H\rho - \rho H .}{time_diff}
We call (\ref{time_diff}) the equation of motion.

\section{Condensation and Classical System}

We propose the concept of condensation which is 
expected in a system with many degrees of freedom. 

\medskip
\noindent
{\bf Definition 4  (Condensation of a system):} 
{\it System $T$ is condensed in the period between 
$\tau_1$ and $\tau_2$, 
if the following conditions are satisfied: 
{\rm (i)} The Hilbert space of the system $T$ is divided into 
a finite or an infinite number of subspaces, and 
{\rm (ii)} if an i-operator describing the system belongs to 
one of the subspaces at a time in this period, 
then the i-operator belongs to the same subspace during the period.} 
\medskip

We call each of the above subspaces the subspace of condensation, 
and call the period $\Delta \tau = \tau_2 - \tau_1$ the period 
of condensation. 
An i-operator belonging to a subspace of condensation 
brings 0 if it operates on any vector out of the subspace. 
If a condensed system is described by an i-operator 
$\rho = p_1 \rho_1 + p_2 \rho_2$ with $\rho_1$ and $\rho_2$ 
which belong to different subspaces of condensation, then 
probabilities $p_1$ and $p_2$ do not change in the period 
($\tau_1 < t < \tau_2$) of condensation. 

A system which is condensed is supposed to be of 
a large number of degrees of freedom. 
Especially, in a system with an infinite number of 
degrees of freedom, the Hilbert space is eternally  divided 
into subspaces of condensation, so that there exists no local 
operator which transforms a vector in a subspace into 
a vector in another subspace. 
This is the case of $\tau_1 \rightarrow - \infty$ and 
$\tau_2 \rightarrow + \infty$. 
This phenomenon corresponds to the spontaneous symmetry 
breaking in a conventional quantum theory for a system with 
an infinite number of degrees of freedom. 
The condensation is a property of the Hamiltonian of the system, 
or the time-development unitary transformation. 

We label a subspace of condensation by $m$, and denote it 
as $\calH_T^m[\tau_1, \tau_2]$ or simply $\calH_T^m$. 
Then the total Hilbert space is written as 
$\calH_T = \oplus_m \calH_T^m$. 
The label $m$ takes discrete values if the number of the subspaces 
is countable, and continuous values if it is uncountable. 

By defining the projection operator to Hilbert space $\calH_T^m$ 
as $P_T^m$, we have the identity 
\algn{\rho_T(t) = \sum_m P_T^m \rho_T(t) P_T^m , \ \ (\tau_1 < t < \tau_2)}{condense}
where, if $m$ is continuous, 
the symbol of summation means integration. 
If the system $T$ is described by $\rho_T(t)$, 
the probability that the system is described by an i-operator 
in the subspace with label $m$, is given as 
\algn{p_T^m = \mathrm{tr}\{ P_T^m \rho_T(t) P_T^m \} \ \ (\tau_1 < t < \tau_2)}{p_T_m}
from (\ref{condense}) and Postulate 4. 
The probability $p_T^m$ is constant in the period of 
$\tau_1 < t < \tau_2$. 
We can know the label $m$ by a measurement as mentioned later. 
If we know a value of the label $m$, we can exclude 
the possibilities other than $m$ in (\ref{condense}). 
Then we can also describe the system by only the term with $m$ as 
\algn{\rho_T^m(t) = 
\frac{P_T^m \rho_T(t) P_T^m}{\mathrm{tr}\{P_T^m \rho_T(t) P_T^m\}} , \ \ (\tau_1 < t < \tau_2)}{rho_T_m}
where the denominator is introduced for the normalization of 
$\mathrm{tr}\{\rho_T^m\} = 1$. 
Since (\ref{condense}) is rewritten as 
\algn{\rho_T(t) = \sum_m p_T^m \rho_T^m(t) , \ \ (\tau_1 < t < \tau_2)}{condense_p}
$\rho_T(t)$ is a common expansion of $\rho_T^m(t)$ 
with arbitrary $m$.  
The system is described by $\rho_T^m(t)$ 
if we include the information of the label $m$, 
while it is simultaneously described still by $\rho_T(t)$ 
which lacks the information.  

We define a classical system based on the concept of condensation 
as follows. 

\medskip
\noindent
{\bf Definition 5  (Classical system):} 
{\it Let a condensed system be described by an i-operator 
which belongs to a subspace labeled by $m$. 
We observe the value of $m$ in a period fairly longer than 
the period of condensation. 
The i-operator changes the belonging subspace 
slowly and successively in the observed period. 
If we do not concern ourselves with anything but the time 
development of the label $m$, then the system is defined 
to be a classical system.} 
\medskip

Classical mechanics is supposed to describe the motion or 
the time development of the label $m$ 
with deleting or averaging the other degrees of freedom. 

Now we refer to a famous subject known as Schr\"odinger's cat. 
The typical situation is as follows: 
Let an alive cat be in an nontransparent box with a lid 
together with an appropriate amount of radium, 
a detector of $\alpha$-particles, and a device scattering 
hydrocyanic acid by trigger of $\alpha$-particles. 
After a while, say $\Delta t$, 
the cat dies owing to hydrocyanic acid 
if an $\alpha$-particle is detected, while it lives if not. 
Let the probabilities of ejecting and not ejecting 
an $\alpha$-particle in the period $\Delta t$ be respectively 
$p^+$ and $p^-$ ($p^+ + p^- = 1$). 

The cat, which we call system $T$, is of a large number of degrees 
of freedom and the problem of "dead or alive" of the cat is 
explained in terms of condensation. 
That the cat is alive is that system $T$ is condensed in the 
subspace with label $m=+$. 
Similarly that the cat is dead is that system $T$ is condensed in the subspace with label $m=-$. 
We denote i-operators describing them as $\rho_T^+$ and 
$\rho_T^-$, respectively. 

After the period $\Delta t$, 
if one finds whether the cat is dead or alive by looking into the box 
and includes the fact as information, 
then the cat is described by only one of $\rho_T^+$ and 
$\rho_T^-$ depending on the fact. 
On the other hand, $\rho_T \equiv p^+ \rho_T^+ + p^- \rho_T^-$ 
is a common expansion of $\rho_T^+$ and $\rho_T^-$, 
and always describes the cat. 
$\rho_T$ is the i-operator where one did not look into the box 
or did look but did not include the fact as information. 
This just means that the probabilities of the cat alive and dead 
are $p^+$ and $p^-$, respectively, when one neglects or 
does not include the information of the fact about the cat. 
Hence the double descriptions by $\rho_T$ and one of 
$\rho_T^+$ and $\rho_T^-$ are not in contradiction. 

Since the cat is condensed for a long time, 
even a part of $\rho_T^+$ ($\rho_T^-$) does not change 
into an i-operator in the subspace of $m=-$ ($m=+$). 
Further there is no i-operator corresponding to 
a vector superposed of vectors in the different subspaces 
unlike the conventional quantum mechanics. 

The reason why Schr\"odinger's cat is paradoxical 
in the conventional quantum mechanics is that 
a condensed system is not distinguished from 
an non-condensed system and also a multiple description 
of a system is not considered.

\section{Composite System}

Consider two systems $S$ and $T$, whose Hilbert spaces are 
$\calH_S$ and $\calH_T$, respectively. 
We denote the composite system consisting of $S$ and $T$ 
by $S + T$, and the Hilbert space by 
$\calH_{S + T}$=$\calH_S \otimes \calH_T$. 
We require the following postulate for consistency. 

\medskip
\noindent
{\bf Postulate 6 (i-operator for a composite system):} 
{\it If i-operators $\rho_S$ and $\rho_T$ describe 
systems $S$ and $T$, respectively, then 
the tensor product $\rho_S \otimes \rho_T$ is an i-operator 
which describes the composite system $S + T$.} 
\medskip

If an i-operator is of the tensor product form, 
$\rho_S \otimes \rho_T$, then it is called 
to be separated into $\rho_S$ and $\rho_T$. 
We further require the following postulate about the separation 
of i-operators. 

\medskip
\noindent
{\bf Postulate 7 (Separation of an i-operator):} 
{\it If systems $S$ and $T$ does not interact with each other, 
and an i-operator $\rho_T$ describes the system $T$, then 
there exists at least one i-operator $\rho_S$ describing $S$ 
such that $\rho_S \otimes \rho_T$ describes 
the composite system $S + T$.} 
\medskip

In Postulate 7, 
$\rho_S$ is called an i-operator corresponding to $\rho_T$. 

\section{Condensation of Composite System}

Consider a composite system $S + T$ consisting of systems 
$S$ and $T$. 
Let the system $T$ be of a large number of degrees of freedom, 
and is condensed with a sufficiently long period of condensation 
if it is isolated. 
Let systems $S$ and $T$ interact with each other 
only in the period of $t_1 < t <t_2$ so that
the condensation is dissolved in the period.  

For $t<t_1$, system $T$ does not interact with system $S$. 
Due to the continuity of i-operators, the composite system 
$S + T$ is described by a separated i-operator as 
\algn{\rho_{S+T}(t_1) = \rho_S(t_1) \otimes \rho_T(t_1).}{rho_ST_t1}
For $t_1<t<t_2$, where the condensation is dissolved, 
the system $S+T$ develops by the unitary operator 
$U_{S+T}(t_1, t_2)$ and is described by 
\algn{\rho_{S+T}(t_2) = 
U_{S+T}(t_1, t_2) \rho_{S+T}(t_1) U_{S+T}^{\dagger}(t_1, t_2).}{rho_ST_t2}
The Hamiltonian $H_{S+T}$ defined by 
$U_{S+T}(t_1, t_2) = \exp \{-i(t_2 - t_1)H_{S+T}/\hbar\}$ 
is supposed to be of the following form: 
\algn{H_{S+T} = H_S + H_T + H_{\rm int},}{Ham_int}
where $H_S$ and $H_T$ are Hamiltonians of systems 
$S$ and $T$, respectively, and 
$H_{\rm int}$ is an interaction term. 
$H_{\rm int}$ depends on time such that 
it vanishes for $t<t_1$ and $t>t_2$. 

Since system $T$ is condensed again for $t>t_2$, 
there exists a set $\{ \rho_T^m(t_2) \}$ of i-operators 
for each subspace of condensation such that 
$\rho_{S+T}(t_2)$ is expanded as 
\algn{\rho_{S+T}(t_2) = 
\sum_m p^m \rho_S^m(t_2) \otimes \rho_T^m(t_2),}{rho_ST_t2_expand}
where $p^m \rho_S^m(t_2)$ is 
the operator-valued expansion coefficient for $\rho_T^m(t_2)$. 
The factor $p^m$ in $p^m \rho_S^m(t_2)$ is determined 
as the remnant for $\rho_S^m(t_2)$ whose trace is unity. 
Due to Postulate 4, (\ref{rho_ST_t2_expand}) means 
that the composite system $S+T$ is described by 
$\rho_S^m(t_2) \otimes \rho_T^m(t_2)$
with probability $p^m$. 
Hence all the factors $\{ p^m \}$ are positive. 
Intuitively speaking, the relation between i-operators 
$\rho_S^m(t_2)$ and $\rho_T^m(t_2)$ is as follows: 
when system $T$ is confined in a subspace of condensation, 
system $S$ which interacts with system $T$ is forced 
in a restricted region of the Hilbert space of $S$. 

If, in addition to the i-operator $\rho_{S+T}(t_2)$, 
we have the information that the label of the system $T$ 
is $m$, then we can describe the composite system $S+T$ 
by $\rho_S^m(t_2) \otimes \rho_T^m(t_2)$. 
Since system $S$ is isolated for $t > t_2$, the i-operator 
\algn{\rho_S^m(t_2)}{rho_S_m_t2}
describes the system $S$. 
The same argument stands for any label $m$ 
which is given as information. 
If we consider system $S$ and do not include the information 
of system $T$ having label $m$, then the i-operator describing 
system $S$ is 
\algn{{\tilde \rho_S}(t_2) \equiv \sum_m p^m \rho_S^m(t_2).}{rho_S_t2_tilde}

By using the time-development unitary operator 
$U_S(t_1, t_2) \equiv \exp\{-iH_S(t_2 - t_1)\}$, 
we define i-operator $\rho_S^m(t_1)$ as 
\algn{\rho_S^m(t_2) = U_S(t_1, t_2) \rho_S^m(t_1) U_S^{\dagger}(t_1, t_2).}{rho_S_m_t1}
That is, $\rho_S^m(t_1)$ is an i-operator which is 
developed in reverse time from $\rho_S^m(t_2)$ 
when system $S$ does not interact with system $T$. 
By the same unitary operator, i-operator (\ref{rho_S_t2_tilde}) 
develops in reverse time to the i-operator 
\algn{{\tilde \rho_S}(t_1) \equiv \sum_m p^m \rho_S^m(t_1).}{rho_S_t1_tilde}
The i-operator ${\tilde \rho_S}(t_1)$ is not the same 
as $\rho_S(t_1)$ in (\ref{rho_ST_t1}), but 
it describes the system $S$ when we do not include 
the information that system $S$ interacts with system $T$. 
We usually know only the i-operator ${\tilde \rho_S}(t_1)$ 
from experiment or calculation at $t_2$, 
and not $\rho_S(t_1)$ since the time-development calculation 
for the composite system $S+T$ by using $U_{S+T}(t_1, t_2)$ 
is very difficult. 

Returning to (\ref{rho_S_t2_tilde}), it means that 
system $T$ determines a set $\{ \rho_S^m(t_2) \}$ of i-operators 
for system $S$. 
For example, we consider a single spin ${\vv s} = (s_x, s_y, s_z)$ 
with magnitude $\frac12$ as system $S$. 
If system $T$ gives $\{ \ket{\uparrow}\bra{\uparrow}, \ket{\downarrow}\bra{\downarrow} \}$ as $\{ \rho_S^m(t_2) \}$, 
then it distinguishes spin orientations in the $z$-direction, 
where $\ket{\uparrow}$ and $\ket{\downarrow}$ are 
eigenvectors of $s_z$. 
On the other hand, 
if system $T$ gives $\{ \ket{\rightarrow}\bra{\rightarrow}, \ket{\leftarrow}\bra{\leftarrow} \}$, 
then it distinguishes spin orientations in the $x$-direction, 
where $\ket{\rightarrow}$ and $\ket{\leftarrow}$ are 
eigenvectors of $s_x$. 
Thus, the system $T$, which is condensed by system $S$ 
as a trigger, has a specified quantization axis. 
In contrast, the system $S$ does not distinguish between 
the quantization axes; e.~g., 
$\frac12 ( \ket{\uparrow}\bra{\uparrow} + \ket{\downarrow}\bra{\downarrow} )$ and 
$\frac12 ( \ket{\rightarrow}\bra{\rightarrow} + \ket{\leftarrow}\bra{\leftarrow} )$ are completely the same. 

The set $\{ p^m \}$ of probabilities in (\ref{rho_S_t2_tilde}) 
is not determined just by system $T$. 
We argue it in the next section of measurement.

\section{Measurement}

Based on the preceding section, we define the {\it measurement} 
as follows: 

\medskip
\noindent
{\bf Definition 6 (Measurement):} 
{\it Let systems $S$ and $T$ form a composite system, 
where both interact with each other only in a finite period. 
The system $T$ is condensed with a sufficiently large 
period of condensation except for the interaction period, and 
the condensation is dissolved only in the interaction period. 
Then, we call the system $S$ the object of measurement, 
the system $T$ the apparatus of measurement, and 
the composite system $S + T$ the measurement system. 
We say that the object $S$ of measurement is measured 
by the apparatus $T$ of measurement. 
After the measurement, 
if we describe the apparatus $T$ of measurement by 
an i-operator in a subspace of condensation with label $m$, 
then we call $m$ the value of the scale 
in the apparatus $T$ of measurement.} 
\medskip

A measurement is a phenomenon between two systems 
which interact with each other. 
The phenomenon is completely objective irrespective of 
the existence or the consciousness of any observer. 
If such a phenomenon naturally takes place without 
human concern, we also refer it as a measurement. 
Therefore, there is no additional postulate to explain 
measurements. 

The system $S+T$ in the preceding section is 
a measurement system. 
The system $S$ at $t_1$ is described by $\rho_S(t_1)$ 
as seen in (\ref{rho_ST_t1}). 
Then the system $S$ interacts with the system $T$ 
in the period of $t_1 < t < t_2$, and we have 
the value $m$ of the scale in the apparatus $T$ of measurement. 
Hence we can describe the system $S$ at $t_2$ 
by $\rho_S^m(t_2)$ in (\ref{rho_S_m_t2}). 
In the measurement process, $\rho_S(t_1)$ does not become 
$\rho_S^m(t_2)$ by any unitary transformation. 
Although the change from $\rho_S(t_1)$ to $\rho_S^m(t_2)$ 
corresponds to a collapse of a wave function 
in the conventional quantum mechanics, 
it is not a jump in the time development. 
We just selected an i-operator 
among multiple i-operators describing the system $S$ 
for our purpose after the measurement. 
In fact we may select $\rho_{S+T}(t_2)$ in (\ref{rho_ST_t2}) 
to describe the composite system $S+T$ at $t_2$. 
However this i-operator involves quantities of $T$ as well as 
those of $S$, so that it is difficult to extract useful information 
about the system $S$. 
A typical observer needs the value of the scale and its probability 
in the apparatus $T$. 
And, after the measurement, he needs an i-operator describing 
only the object $S$ of measurement. 
Then, if he thinks reasonably, he describes the system by $\rho_S$ 
before measurement, and describes it by $\rho_S^m$ after 
measurement if he has the information that the value of the scale 
is $m$. 
Since both $\rho_S$ and $\rho_S^m$ are separately continuous 
with respect to time, the measurement introduce 
no discontinuity. 

We now determine the set $\{ p^m \}$ of probabilities 
in ${\tilde \rho_S}(t_2)$ of (\ref{rho_S_t2_tilde}), 
which describes the system $S$ after the measurement. 
To be precise, we adopt $\{ p^m \}$ rather than determine it. 
At first, we consider the case that we know the i-operator 
(\ref{rho_ST_t1}) describing the measurement system $S+T$ 
at $t_1$. 
Also let it be possible to calculate (\ref{rho_ST_t2}) 
by the time-development unitary operator $U_{S+T}(t_1, t_2)$. 
Further let it be possible to expand 
the obtained i-operator in the form of 
(\ref{rho_ST_t2_expand}). 
In this case, we may adopt $\{ p^m \}$, 
which has been determined in the expansion coefficients, 
as a set of probabilities. 
This i-operator $\rho_{S+T}(t_2)$ describes the composite 
system at $t_2$ in the case that we know $\rho_{S+T}(t_1)$ 
as information before the measurement. 
However, it is not necessarily possible to know $\rho_{S+T}(t_1)$ 
or to calculate $\rho_{S+T}(t_2)$. 

Another way of adopting $\{ p^m \}$ is found, 
if we prepare $N$ equivalent measurement systems. 
The number $N$ need not but may be large. 
We perform the equivalent measurement 
for each measurement system. 
Then, as $p^m$ for each value $m$ of the scale, 
we adopt the ratio of the obtained number of 
the value $m$ against the total number $N$ of 
the experiments. 
This choice of $\{ p^m \}$ determines an i-operator 
(\ref{rho_S_t2_tilde}) which describes the system $S$. 
This is the i-operator which includes the information 
from the specific $N$ experiments. 

We now perform another set of equivalent $N$ experiments. 
Then we have other values for $\{ p^m \}$, since 
the number obtaining the value $m$ is generally different 
from the previous. 
We hence adopt another set of values for $\{ p^m \}$. 
Thus the system $S$ is described by a different i-operator 
for each set of experiments. 
There is no contradiction between these descriptions, 
since each description just includes different information 
about the system $S$. 
As an extreme case, we can specify arbitrary values for 
$\{ p^m \}$ satisfying $\sum_m p^m =1$ 
without performing experiments. 
Then we can say that (\ref{rho_S_t2_tilde}) with 
the values for $\{ p^m \}$ also describes the system $S$. 
However the i-operator does not include any information from 
any experiments and hence is not useful. 
Thus it is important 
not only that the i-operator describes the system 
but also that it includes information which we need. 

We compare the results, when 
we repeat many sets of $N$ experiments. 
Although the sets of the obtained values for $\{ p^m \}$ 
are generally different from each other, 
we may usually have the following physical expectation. 
That is, many sets of the values for $\{ p^m \}$ 
may be close to each other. 
As a useful choice in the case, we adopt 
the i-operator (\ref{rho_S_t2_tilde}) with $\{ p^m \}$ 
where each $p^m$ is the average for the sets of experiments. 
On the other hand, we do not exclude the possibility that 
the values for $\{ p^m \}$ depend strongly on 
each set of experiments. 
In such cases, we do not predict the results for 
the next set of experiments. 

From (\ref{rho_S_t2_tilde}), ${\tilde \rho}_S(t_2)$ is 
a common expansion of $\{ \rho_S^m(t_2) \}$ from Theorem 3. 
We denote the contracting operator for each $m$ as $K_S^m$;  
i.~e. 
$\rho_S^m(t_2) = K_S^m {\tilde \rho}_S(t_2) (K_S^m)^{\dagger}$. 
Then we have the following equation: 
\algn{{\tilde \rho}_S(t_2) = \sum_m p^m K_S^m {\tilde \rho}_S(t_2) (K_S^m)^{\dagger}.}{condense_S}
By defining $M_S^m \equiv \sqrt{p_T^m} K_S^m$ 
$(m=1, 2, \cdots)$, this equation becomes 
\algn{{\tilde \rho}_S(t_2) = \sum_m  M_S^m {\tilde \rho}_S(t_2) (M_S^m)^{\dagger}.}{condense_S_M}
Then we have 
\algn{p^m = \mathrm{tr}\{ M_S^m {\tilde \rho}_S(t_2) (M_S^m)^{\dagger} \}.}{probability_S}

In a general measurement system, 
$\{ M_S^m \}$ in (\ref{condense_S_M}) depends 
on ${\tilde \rho}_S(t_2)$. 
We define a special class of measurement systems as follows: 

\medskip
\noindent
{\bf Definition 7 (Definitive Measurement):} 
{\it A measurement system $S+T$ where 
{\rm (\ref{condense_S_M})} 
is an identity is a definitive measurement system.} 
\medskip

That is, $\{ M_S^m \}$ is determined independently of 
each ${\tilde \rho}_S(t_2)$ in a definitive measurement system. 
By taking a trace of (\ref{condense_S_M}), we have 
$\mathrm{tr} \{ {\tilde \rho}_S(t_2) \sum_m (M_S^m)^{\dagger} M_S^m \}$ = 1. 
Since this stands for any $\rho_S(t_2)$ in the definitive 
measurement system, we have 
\algn{\sum_m (M_S^m)^{\dagger} M_S^m = I_S.}{condition}
By using $\{ M_S^m \}$, the observable is defined as follows: 

\medskip
\noindent
{\bf Definition 8 (Observable):} 
{\it In a definitive measurement system $S+T$, Hermitian operator 
\algn{F = \sum_m f(m) (M_S^m)^{\dagger} M_S^m}{observable}
with a real function $f(m)$ is an observable of the system $S$ 
for the apparatus $T$ of measurement.} 
\medskip

Although $M_S^m$ is an operator in the Hilbert space $\calH_S$, 
the definition depends on the condensation of the system $T$. 
In general, $f(m)$ is not an eigenvalue of $F$. 
In the special case that $M_S^m$ is the projection operator 
for all $m$, 
the definition of $F$ is the same as that of an observable in 
the conventional quantum mechanics and 
$f(m)$ becomes an eigenvalue of $F$. 

By (\ref{probability_S}) and (\ref{observable}), 
the expectation value of the observable $F$ is written as 
\algn{\bar{F} = \sum_m f(m) \, p^m 
= \mathrm{tr} \{ F \rho_S(t_2) \}.}{expectation}
This reads as follows: 
When observable $F$ of the object $S$ of measurement is 
measured by the apparatus $T$ of measurement, 
the probability that the value of the scale is $m$ is $p^m$. 
Then the value of observable $F$ is $f(m)$. 
If we repeat $N$ times the same measurement 
to $N$ equivalent systems, then 
we physically expect that the average of the obtained values of 
$f(m)$ reaches $\bar{F}$ with increasing $N$.

\section{Information Vector}

As has been defined, an i-operator $\rho$ is pure if it is written as 
\algn{\rho = \ket{\psi} \bra{\psi}}{rho_pure}
with a vector $\ket{\psi}$ in the Hilbert space of the system. 
This expression is not unique, since we can use 
\algn{\ket{\psi'}=e^{i\alpha}\ket{\psi}}{phase_arb}
with arbitrary real number $\alpha$ and write it as 
\algn{\rho = \ket{\psi'} \bra{\psi'}.}{rho_pure_d}

For a pure i-operator, $\rho = \ket{\psi} \bra{\psi}$, 
we can identify it by specifying the vector $\ket{\psi}$, or 
$\ket{\psi'}=e^{i\alpha}\ket{\psi}$. 
Hence it is allowed to use the vector $\ket{\psi}$ 
as an agent for $\rho$. 
In this usage, we call $\ket{\psi}$ the information vector. 
Thus an arbitrariness  of a phase factor is brought out, 
although it is not in the original i-operator. 
The information vector $\ket{\psi}$ is only an agent for 
the corresponding i-operator, and does not represent a {\it state} 
of the system in the conventional quantum mechanics. 
Even if a pure i-operator describes a system, 
another i-operator which is not pure may describe 
the same system simultaneously. 

On the other hand, we can treat an information vector 
as if it is a state vector in the conventional quantum mechanics. 
In particular, we have a new information vector 
by superposing two information vectors. 
An information vector describing the system follows 
the Schr\"{o}dinger equation. 
The probability interpretation stands like the conventional 
quantum mechanics. 
Therefore the information vector covers 
all the same description region of the state vector 
in the conventional quantum mechanics.

\section{Stern-Gerlach Experiment}

We consider the Stern-Gerlach experiment with an apparatus in 
a standard arrangement. 
Two magnets are placed in a way that the north pole of a magnet 
and the south pole of the other magnet face each other in the $z$-direction. 
The shape of one magnet is acute and that of the other is plane 
so as to yield an inhomogeneous magnetic field. 
Let silver atoms be ejected and 
let them travel through the magnetic field. 
Behind the magnets a screen is placed to stop the silver atoms. 
A silver atom has a spin $\vv{s}$ of magnitude $s=\frac{1}{2}$ 
which comes from the outermost electron. 
The ejected atoms are very dilute so that the problem is of a single atom. 
After passing through the magnetic field, the atom is deflected 
positively or negatively in the $z$-direction and 
makes a spot on one of two places of the screen. 
This is because the atom is attracted to the negative direction 
if the $z$-component of the spin is $+\frac{1}{2}$ and 
to the positive direction if it is $-\frac{1}{2}$ 
in the inhomogeneous magnetic field. 
The deflection becomes opposite depending on the arrangement 
and the forms of the magnets. 
We examine this phenomenon in terms of i-operators 
as follows. 

When a silver atom is ejected, it is not affected by 
the distant magnets. 
The atom reaches and passes through the magnet, so that 
it interacts with the magnets only in a finite period $t_1 < t < t_2$. 
After then, the atom is isolated from the magnets again. 
When the atom does not interact with the magnets, 
the spin degree of freedom is also independent of 
the orbital degree of freedom in the atom. 
At $t=t_1$, let $\rho_{\mathrm{mag}}(t_1)$ be an i-operator 
describing the magnets, $\rho_{\mathrm{orb}}(t_1)$ be 
an i-operator describing the orbital degree of freedom, 
$\rho_S(t_1)$ be an i-operator describing 
the spin degree of freedom. 
Then the total system is described by the i-operator 
$\rho_S(t_1) \otimes \rho_{\mathrm{orb}}(t_1) \otimes \rho_{\mathrm{mag}}(t_1)$. 
As for the spin degree of freedom, we denote the eigenvector 
for eigenvalue $+\frac{1}{2}$ of $s_z$ as $\ket{\uparrow}$, 
and the eigenvector for $-\frac{1}{2}$ as $\ket{\downarrow}$. 
Then the corresponding pure i-operators are 
$\rho_S^{\uparrow} \equiv \ket{\uparrow} \bra{\uparrow}$ and 
$\rho_S^{\downarrow} \equiv \ket{\downarrow} \bra{\downarrow}$, respectively. 
Since we have no information about the spin direction of 
the ejected atom, we describe the spin by the i-operator 
$\rho_S(t_1)$ = 
$\frac{1}{2} (\rho_S^{\uparrow} + \rho_S^{\downarrow})$. 

Here we regard the total system as a composite system 
of systems $S$ and $T$, where the system $S$ is only of 
the spin degree of freedom in the atom, and 
the system $T$ is further a composite system of the magnets 
and the orbital degree of freedom of the atom. 
The latter is described by the i-operator $\rho_T(t_1)$ $\equiv$ 
$\rho_{\mathrm{orb}}(t_1) \otimes \rho_{\mathrm{mag}}(t_1)$. 
Then the total system at $t = t_1$ is described by i-operator 
$\rho_{S+T}(t_1) = \rho_S(t_1) \otimes \rho_T(t_1)$, 
which is of the same form as (\ref{rho_ST_t1}). 

The system $T$ is condensed for $t<t_1$ and $t>t_2$, 
in which periods it does not interact with the system $S$. 
Actually, since the magnets are macroscopic, 
it is supposed that the weight or probability for the part with  
each value of the momentum of the magnets 
does not change with time in a total i-operator describing 
the system $T$. 
The orbital degree of freedom of the atom does not affect 
the condensation of the magnets and the system $T$ is 
also condensed. 
Here we define that the value $m$ of the scale of measurement 
takes $-1$, 0 or 1 according to the sign of the $z$-component 
of the momentum of the system $T$. 
For $t<t_1$, the magnets stands still and then the system $T$ 
is condensed with $m=0$. 
For $t_1 < t < t_2$, the system $T$ interacts with 
the system $S$, so that the condensation is dissolved and 
then the system $T$ obtains a different value of the momentum. 
Since the momentum difference is quite small, 
we cannot detect it in reality. 
However, since the total momentum of the magnets and the atom  
is conserved, the change of the momentum of magnets 
is reflected in the atomic orbital for $t>t_2$. 
Therefore we read the value $m$ 
as $-1$ for an orbital deflected above, 
0 for an straight orbital, and 1 for an orbital deflected below. 

We consider phenomenologically the time development of 
the system $S+T$ for $t_1 < t < t_2$, in which period 
the system is dissolved from the condensation. 
Typical i-operators for the system $T$ are 
$\rho_T^0 = \ket{\psi_T^0} \bra{\psi_T^0}$, 
$\rho_T^+ = \ket{\psi_T^+} \bra{\psi_T^+}$ and 
$\rho_T^- = \ket{\psi_T^-} \bra{\psi_T^-}$: 
$\rho_T^0$ represents the motion that the atomic orbital is 
straight and the $z$-component of the momentum of 
the magnets is zero; 
$\rho_T^+$ ($\rho_T^-$)  represents the motion that 
the atomic orbital is deflected upward (downward) and 
the $z$-component of the momentum of the magnets 
is negative (positive). 
Here we define a pseudo spin $\vv{R} = (R_x, R_y, R_z)$ 
of magnitude unity 
such that $\ket{\psi_T^-}$, $\ket{\psi_T^0}$ and 
$\ket{\psi_T^+}$ are eigenvectors belonging to eigenvalues, 
$-1$, 0 and 1, respectively, of $R_z$. 

The time development of an i-operator in the period of 
$t_1 < t < t_2$ is given by the Hamiltonian $H$ 
for the system $S+T$ or by the unitary operator 
$U(t_1, t_2) = \exp(-iH (t_2 - t_1) /\hbar)$. 
Here we express this time development by the following 
phenomenological unitary operator $U$ instead of $U(t_1, t_2)$: 
\algn{
U = \frac{1}{4} [ (1 - 2s_z) \otimes {\tilde R}_+ 
+ (1 + 2s_z) \otimes {\tilde R}_- ] ,}{stern_U}
with 
\algn{
{\tilde R}_{\pm} \equiv \sqrt{2}(R_z \pm 1) R_x (R_z \pm 1) 
+ R_z (R_z \mp 1) .}{stern_R}

Since the atomic orbital is straight before the interaction, 
the system $S+T$ at $t = t_1$ is described by i-operator 
\algn{\rho_{S+T}(t_1) = \frac{1}{2} 
( \rho_S^{\uparrow} + \rho_S^{\downarrow}) 
\otimes \rho_T^0 .
}{stern_t1}
By straightforward calculation, 
the i-operator develops to the following i-operator at $t = t_2$: 
\algn{
\rho_{S+T}(t_2) = U \rho_{S+T}(t_1) U^{\dagger} 
= \frac{1}{2} ( \rho_S^{\uparrow}\otimes \rho_T^- 
+ \rho_S^{\downarrow}\otimes \rho_T^+ ) .}{stern_t2}
Hence the system $S+T$ at $t = t_2$ is described by 
$\rho_{S+T}(t_2)$. 
This i-operator means that the system is described by 
$\rho_S^{\uparrow}\otimes \rho_T^-$ with probability 
$\frac{1}{2}$ and by $\rho_S^{\downarrow}\otimes \rho_T^+$ 
with probability $\frac{1}{2}$. 
For $t>t_2$, since the system is condensed again, 
$\rho_{S+T}(t_2)$ keeps the same form as 
(\ref{stern_t2}): i.~e. 
$\rho_T^-(t)$ ($\rho_T^+(t)$) develops within the subspace 
of $m=-$ ($m=+$) without mixing. 

We suppose to have the information that the atom reached 
the screen and a spot appeared at a downward deflected position. 
It means that we have $m=-$ for the value of the scale 
in the system $T$. 
The probability that this occurs is $\frac{1}{2}$ 
from (\ref{stern_t2}). 
If we include this information, we describe the total system 
by $\rho_S^{\uparrow}\otimes \rho_T^-$. 
Owing to the separated form of the i-operator, 
the system $S$ is described by i-operator $\rho_S^{\uparrow}$. 
Similarly, we suppose to have the information that the atom reached 
the screen and a spot appeared at an upward deflected position, 
and we suppose to include the information. 
Then we describe the system $S$ by i-operator 
$\rho_S^{\downarrow}$. 

The conventional quantum mechanics may explain 
the situation that the value of the scale is $m=-$ as follows: 
the density operator (\ref{stern_t2}) discontinuously 
changes into $\rho_S^{\uparrow}\otimes \rho_T^-$. 
In contrast, the present reconstructed quantum mechanics 
produces no discontinuity in any i-operators. 
For $t>t_2$, the total system is described 
by the i-operator (\ref{stern_t2}) as well as 
$\rho_S^{\uparrow}\otimes \rho_T^-$. 
These two i-operators are continuous 
for all period including $t \le t_2$. 
An observer usually prefers the information of $m=-$, and 
describes the system by $\rho_S^{\uparrow}\otimes \rho_T^-$. 
On the other hand, an observer may not care the value $m$, 
or may not have the information of $m=-$. 
Then the observer may describe the same system by 
i-operator (\ref{stern_t2}). 
If we pay attention only to the system $S$, we say the followings: 
The system $S$ is described by $\rho_S^{\uparrow}$ 
if the information of $m=-$ is included, and by 
$\frac{1}{2} (\rho_S^{\uparrow} + \rho_S^{\downarrow})$
if the information is not included.

\section{Electron through Screen with Slits}

We examine a slit experiment where an electron is ejected to 
a solid thin screen that has typically two slits cut into it. 
Behind the slit screen, another detection screen is also 
set up to record what comes. 
This is performed as a real experiment~\cite{tonomura}, 
and is regarded as an interference experiment of an electronic 
wave function in the conventional quantum mechanics. 
Hereafter we consider only the orbital degree of freedom, 
which affects experimental results, and 
do not pay attention to the spin degree of freedom. 

The electron reached the slit screen is absorbed or reflected 
by the material of the slit screen with a large probability, and 
cannot go through it. 
Only in a small probability, say $p$, the electron 
reaches the detection screen. 
Then the electron interacts with the slit screen for 
$t_a < t < t_b$. 
Thus the electron from the ejection time to the time reached to 
the slit screen is supposed to be described by the i-operator 
\algn{\rho(t) = p \rho_a(t) 
+ (1-p) \rho_\mathrm{abs}(t) , 
\ \ (t < t_a)}{rho_before}
where $\rho_a(t) \equiv \ket{\psi_a(t)} \bra{\psi_a(t)}$ 
is an i-operator representing the passage, 
and $\rho_\mathrm{abs}(t)$ is an i-operator representing 
the non-passage. 
Then $p$ is the probability that the electron passes through the slits. 
By denoting an i-operator describing the slit screen as 
$\rho_\mathrm{slit}(t)$, the total system is described by 
a separated i-operator $\rho(t) \otimes \rho_\mathrm{slit}(t)$ 
$(t < t_a)$. 
After $t_a$, the electron interacts with the slit screen, 
so that the i-operator develops with time into an unseparated 
form. 

Now we suppose to be interested in phenomena for $t>t_b$ 
only when the electron passes through the slit screen. 
In this case, it is not convenient to use the i-operator 
which is time-developed from $\rho(t)$ in (\ref{rho_before}). 
In fact $\rho(t)$ is not separated into 
an electronic factor and a slit screen factor, and further 
it even describes the unnecessary possibility that 
the electron is absorbed in the slit screen to vanish. 
We have examined only the case that the electron 
passes through the slit screen, and the passage is judged 
by a spot on the detection screen. 
In the case that we found a spot, 
the electron clearly passed through the slit screen. 
Hence it is reasonable to describe the electron by a pure i-operator 
\algn{\rho_b(t) = \ket{\psi_b(t)} \bra{\psi_b(t)}. 
\ \ (t > t_b)}{rho_after}
The total system is then described by an i-operator in the form of 
$\rho_b(t) \otimes \rho'_\mathrm{slit}(t)$. 

Here we assume that $\rho_a(t)$, a part of 
(\ref{rho_before}), continues to $\rho_b(t)$ in 
(\ref{rho_after}) with time. 
That is, we assume the existence of a pure i-operator 
$\rho_\mathrm{int}(t) \equiv \ket{\psi_{\rm int}(t)} \bra{\psi_{\rm int}(t)}$ 
such that 
\algn{\tilde{\rho}(t) = 
\left\{
\begin{array}{ll}
\rho_a(t) & (t < t_a)  \\
\rho_\mathrm{int}(t) & (t_a < t < t_b)  \\
\rho_b(t) & (t > t_b)
\end{array}
\right. 
}{rho_electron}
is continuous with time. 
This assumption seems to be physically allowable, 
although it is not trivially guaranteed. 
In terms of information vectors, we have assumed that 
there exists a continuous vector $\ket{\psi(t)}$ such that 
$\ket{\psi(t)} = \ket{\psi_a(t)}$ $(t < t_a)$, 
$\ket{\psi(t)} = \ket{\psi_{\rm int}(t)}$ $(t_a < t < t_b)$ and 
$\ket{\psi(t)} = \ket{\psi_b(t)}$ $(t > t_b)$, if 
the phase factor is appropriately chosen. 
The interference effect for the {\it wave function} of the electron 
in the conventional quantum mechanics is actually for the information vector $\ket{\psi(t)}$.

\section{Summary}

We have reconstructed quantum mechanics based on 
two central concepts. 
One of them is the multiple description of a physical system. 
There are multiple inequivalent i-operators to simultaneously 
describe a single system, and the different i-operators carry 
different kinds or amounts of information for the same system. 
There is no preferential i-operator which 
plays a specially important role. 
What we can do is to draw information out from 
the i-operator which we have. 
Accordingly we discarded the concept of the state, 
or the wave function, which has played the central role 
in the conventional quantum mechanics. 

The other concept in the reconstructed quantum mechanics 
is the condensation of a system. 
For a condensed system, the Hilbert space decomposes 
into subspaces where the time-development unitary operator 
or the Hamiltonian cannot overcome the boundaries of 
the subspaces. 
Reading the label of the subspace, we can adopt an i-operator 
belonging to the subspace to describe the system. 

The measurement is defined as the phenomenon between 
two systems $S$ and $T$ where $T$ is condensed if it is isolated, 
$S$ and $T$ interact with each other in a finite period, and 
the condensation of $T$ is dissolved in the interaction period. 
In the measurement system $S+T$, 
we have called $T$ the apparatus of measurement and 
$S$ the object of measurement. 
The results of the measurement is the label of the subspace 
of condensation and the i-operator belonging to the subspace. 
We have called the label the value of the scale of measurement. 
By the concepts of the multiple description of a system 
by the i-operators and the concept of the condensation 
of the system, we can explain quantum phenomena including 
the measurement without any discontinuity. 
Especially we need no postulate specific to the measurement.

\section*{Acknowledgment}

I would like to thank Yoshio Ohnuki, Yuki Sugiyama and Shinsaku Kitakado for discussions.

\section*{References}

\end{document}